\documentclass[runningheads]{llncs}
\usepackage[T1]{fontenc}
\usepackage{cite}
\usepackage{amsmath}
\usepackage{booktabs} % For professional tables
\usepackage{multirow}
\usepackage{graphicx,verbatim}
\begin{document}
\title{Ultrasound Report Generation with Multimodal Large Language Models for Standardized Texts}

\author{Peixuan Ge\inst{1,2} \and
Tongkun Su\inst{1} \and
Faqin Lv\inst{3} \and
Baoliang Zhao\inst{1} \and
Peng Zhang\inst{1} \and\\
Chi Hong Wong\inst{4} \and
Liang Yao\inst{1} \and
Yu Sun\inst{1} \and
Zenan Wang\inst{1} \and
Pak Kin Wong\inst{2*} \and
Ying Hu\inst{1*}}
\authorrunning{P. Ge et al.}
\institute{Shenzhen Institutes of Advanced Technology, \and
University of Macau, \and 
Chinese PLA General Hospital, \and
Macau University of Science and Technology \\
\email{px.ge@siat.ac.cn}
}

\maketitle              % typeset the header of the contribution
\begin{abstract}
	Ultrasound (US) report generation is a challenging task due to the variability of US images, operator dependence, and the need for standardized text. Unlike X-ray and CT, US imaging lacks consistent datasets, making automation difficult. In this study, we propose a unified framework for multi-organ and multilingual US report generation, integrating fragment-based multilingual training and leveraging the standardized nature of US reports. By aligning modular text fragments with diverse imaging data and curating a bilingual English-Chinese dataset, the method achieves consistent and clinically accurate text generation across organ sites and languages. Fine-tuning with selective unfreezing of the vision transformer (ViT) further improves text-image alignment. Compared to the previous state-of-the-art KMVE method, our approach achieves relative gains of about 2\% in BLEU scores, approximately 3\% in ROUGE-L, and about 15\% in CIDEr, while significantly reducing errors such as missing or incorrect content. By unifying multi-organ and multi-language report generation into a single, scalable framework, this work demonstrates strong potential for real-world clinical workflows.
% Authors must provide keywords and are not allowed to remove this Keyword section.
%This study explores the use of multimodal large language models (MLLMs) for US report generation, as no prior work has specifically investigated this application. Fine-tuning models from multiple families shows that unfreezing and training the vision transformer (ViT) is critical for handling US image noise, with even simple ViT joint fine-tuning outperforming previous state-of-the-art methods. By leveraging the standardized nature of US reports and curating a translated English-Chinese dataset, Compared to previous state-of-the-art methods, this approach achieves relative gains of about 2\% in BLEU scores, 3\% in ROUGE-L, and 15\% in CIDEr. These models also integrate multi-organ and multi-language report generation into a single framework, offering significant potential for real-world clinical workflows. Code and model weights will be released after review.
\end{abstract}
\section{Introduction}
Radiology report generation requires interpreting complex medical images and producing standardized, clinically accurate text. Ultrasound (US) report generation is particularly challenging due to its operator dependence, high variability, and susceptibility to noise. While most research in automated report generation has focused on X-ray and CT~\cite{x_ray_report1,x_ray_report2,x_ray_report3,llm_x_ray_report,report_gen_review}, where standardized datasets are readily available, US remains underexplored~\cite{breast_us_report,KMVE_us_report} due to the lack of standardized datasets. Existing models struggle to generalize across US images or handle multi-organ reporting tasks.

Large Language Models (LLMs) have demonstrated strong text generation capabilities in medical applications~\cite{llm_in_med_review,llm_x_ray_report}, and the integration of visual encoders in multimodal LLMs (MLLMs) has enabled joint image-text understanding~\cite{Qwen2VL,Internvl2_5,Llava_med, HuatuoGPT_Vision}. General-purpose models such as Intern-VL 2.5~\cite{Internvl2_5} and Qwen2-VL~\cite{Qwen2VL}, as well as medical-specific models like HuatuoGPT-Vision~\cite{HuatuoGPT_Vision} and LLaVA-Med 1.5~\cite{Llava_med}, have demonstrated strong performance in vision-language tasks, particularly in visual question answering, which requires answering specific queries about medical images. In contrast, report generation involves producing a streamlined, direct textual output that summarizes findings from images, a task that poses unique challenges for MLLMs. Applying MLLMs to US report generation remains underexplored due to the inherent noise in US images, variable image quality, and the limitations of existing pretraining domains. Prior work in US report generation has largely focused on specific applications like breast ultrasound~\cite{breast_us_report} or organ-specific tasks~\cite{KMVE_us_report}, using datasets that lack the diversity and standardization needed for robust automation.

To address these challenges, this study introduces a framework for US report generation that combines Vision Transformer (ViT)~\cite{vit} optimization and fragment-based multilingual training. By unfreezing the ViT during supervised fine-tuning (SFT), we enable better alignment with the language model, allowing it to adapt to the noisy and variable nature of US images. Our fragment-based multilingual training leverages the standardized and modular structure of US reports to enable consistent bilingual report generation without additional datasets. Experimental results show that this unified framework outperforms traditional models and baseline MLLMs across BLEU~\cite{BLEU}, ROUGE-L~\cite{ROUGE}, and CIDEr~\cite{CIDEr} metrics, achieving significant improvements in multilingual and organ-specific tasks. Qualitative analysis further demonstrates that the approach produces semantically accurate and clinically relevant reports, closely aligned with expert-generated ground truth. These contributions highlight the potential of our framework to advance automated US report generation and its applicability in real-world clinical workflows. To encourage further research, we will release our codebase and model weights after the review process.

\section{Methods}
%	\caption{
	%		Overview of the proposed framework, including the MLLM architecture, dataset fragment statistics, examples of input ultrasound images, raw and translated reports, the fragment-translation lookup table, and multilingual training prompts. 
	%	}
\subsection{Decoder-Only Multimodal LLMs}
\begin{figure}[htbp]
	\centering
	\includegraphics[width=\textwidth]{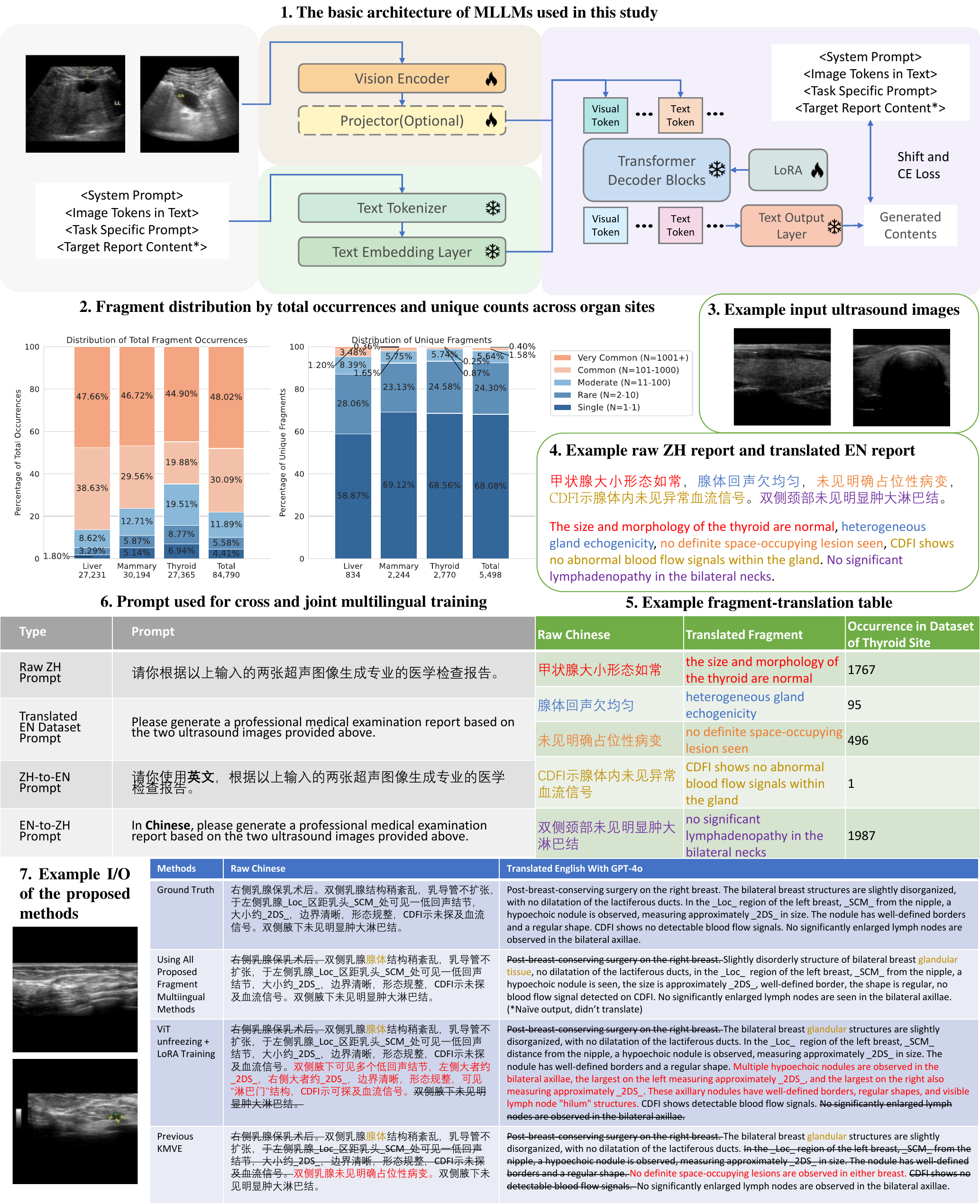}
		\caption{
		(1) The MLLM architecture. *Target report content included only during training to compute the masked causal loss. 
		(2) Fragment distribution by total occurrences and unique counts across organ sites.  
		(3, 4, 5) Example input ultrasound images at thyroid site, raw Chinese and translated English reports, and fragment-translation lookup table mapping Chinese fragments to English equivalents.  
		(6) Multilingual training prompts for joint and cross-language generation.  
		(7) Example inputs and outputs for ground truth, proposed methods and the old method.  
	}
	\label{fig:methods}
\end{figure}
This study employs decoder-only MLLMs to generate standardized, multi-organ US reports in both Chinese and English. The architecture, shown in Figure~\ref{fig:methods}, integrates visual tokens from a vision encoder (ViT) with textual tokens from a tokenizer to jointly perform image-text understanding and generate structured text reports. The input comprises system prompts, image tokens, and user prompts, with target report content included during training using a causal mask to prevent information leakage. During inference, the model generates reports autoregressively without target content in the input.

We evaluate several MLLMs, including general-purpose models like Intern-VL 2.5~\cite{Internvl2_5} and Qwen2VL~\cite{Qwen2VL}, as well as medical-specific models like HuatuoGPT-Vision~\cite{HuatuoGPT_Vision} and LLaVA-Med 1.5~\cite{Llava_med}. These models vary in their vision encoders, tokenization strategies, and pretraining domains. To address the variability and noise of US images, we fine-tune the vision encoder by unfreezing its parameters during SFT. This allows the model to jointly optimize image and text features, addressing the variability and noise of US images, which frozen encoders fail to handle effectively. Additionally, we integrate Low-Rank Adaptation (LoRA)~\cite{LoRA} for efficient fine-tuning of the text decoder, optimizing only specific layers of the Transformer architecture including query, key, value, attention output and MLP projections to reduce computational cost. All models are implemented using the Hugging Face Transformers ecosystem~\cite{transformers}, ensuring compatibility and standardization.

The supervised fine-tuning (SFT) process minimizes the causal loss:
\[
\mathcal{L}_{\text{SFT}} = - \sum_{i=1}^{T} \log p_\theta(x_i \mid x_{<i}),
\]
where \(x_i\) represents the \(i\)-th token in the sequence, \(x_{<i}\) denotes the preceding tokens, and \(\theta\) refers to the model parameters. Sampling strategies such as top-\(k\), top-\(p\), and temperature scaling are applied during inference to improve output diversity. By fine-tuning both the vision encoder and text decoder, our framework achieves better alignment between image and text modalities, significantly improving performance for US report generation compared to traditional frozen-encoder but unfrozen-projector LLaVA~\cite{llava} like approaches.

\subsection{Dataset and Fragment-Based Multilingual Training Pipeline}

The dataset, derived from KMVE~\cite{KMVE_us_report} study, includes three organ scan sites: mammary (3,521 patients), thyroid (2,474 patients), and liver (1,395 patients). Each sample consists of two ultrasound images paired with a standardized Chinese report composed of modular fragments that describe clinical findings like scan quality, disease severity, and anatomical attributes (Figure~\ref{fig:methods}, Section 2). These fragments exhibit a standardized structure, with a few common fragments dominating the dataset, while rarer fragments add diversity. 

To enable bilingual training, we develop a fragment-based translation pipeline. Each report is segmented into fragments using delimiters of commas, semicolons and periods then translated into English. Regex-based checks ensure the preservation of domain-specific terms like "CFDI," and all translated fragments are validated by radiology experts to ensure medical accuracy. Validated fragments are stored in a lookup table (Figure~\ref{fig:methods}, Section 5) across all samples that maps Chinese fragments to their English equivalents. This ensures consistent translation across reports, maintaining high-quality bilingual data for training.

The bilingual dataset is constructed using four training prompts (Figure~\ref{fig:methods}, Section 6): generating Chinese reports from ultrasound images, English reports from ultrasound images, English reports from Chinese queries, and Chinese reports from English queries. These prompts leverage the dataset’s modular structure with tokenization methods like byte-pair encoding (BPE) and SentencePiece to achieve multilingual representation. The fragment-based pipeline further enhances the alignment of multimodal and multilingual features for generating multi-organ, multilingual reports.

\section{Experiments and Results}
All models were trained under the LoRA + ViT unfrozen (including projector) configuration on NVIDIA 4090 GPUs. For inference, a 16GB BF16-capable GPU suffices for 7B models, while training requires a 24GB GPU. Hyperparameters are model-specific, but the best two configurations for 2B and 7B are presented here. Shared hyperparameters for both 2B and 7B models included AdamW optimizer, a cosine learning rate schedule with a warmup ratio of 20\%, weight decay of 0.05, 20 training epochs, LoRA rank 32, alpha 64, and no LoRA bias. Differences included dropout rates (0.1 for 2B, 0.2 for 7B), learning rates (0.0001 for 2B, 0.00005 for 7B), and maximum gradient norms (2.0 for 2B, 1.0 for 7B). An effective batch size of 64 was used through gradient accumulation. 

Evaluation metrics included BLEU~\cite{BLEU} (B1, B4) for n-gram overlap, with BLEU-1 focusing on unigrams and BLEU-4 on up to 4-grams; ROUGE-L~\cite{ROUGE} (RL) for recall via common subsequence; CIDEr~\cite{CIDEr} (C) for contextual relevance with weighted n-grams; BERTScore F1 (BE-ZhF for Chinese , BE-MF for multilingual)~\cite{BERT} for semantic similarity; and Matching Keyword F1 (MKF1) for clinical relevance based on organ-specific keywords such as "thyroid", derived from KMVE~\cite{KMVE_us_report}. Since KMVE did not release formal keyword list, variations arise from differences in interpretation from the text. KMVE reports per-organ metrics but lacks overall results across all three organ sites, as their models are organ-specific. To enable comparison, we reproduced their results, achieving slightly higher values than reported in the original paper for consistency. Per-organ metrics compare MLLMs with traditional models, while overall metrics assess improvements across all sites.

\subsection{Comparison of Traditional Methods and MLLMs}
\begin{table}[ht]
	\caption{Comparison of methods across three organ sites. Results include traditional methods and MLLMs with LoRA+ViT Unfreeze SFT. *Results for DeltaNet~\cite{DeltaNet} and R2GenRL~\cite{R2GenRL} are from KMVE~\cite{KMVE_us_report}, which did not evaluate CIDEr for these methods. The  KMVE results were reproduced in this study to enable comparison with MLLMs.}.
	\label{tab:combined_metrics_table}
	\centering
    \begin{tabular}{l|cccc}
		\toprule
		\multirow{2}{*}{\textbf{Method}} & \multicolumn{4}{c}{\textbf{Metrics (Liver/Mammary/Thyroid)}} \\
		\cmidrule{2-5}
		& \textbf{B1} & \textbf{B4} & \textbf{RL} & \textbf{C} \\
		\midrule
		DeltaNet* & 0.87/0.72/0.61 & 0.81/0.61/\textbf{0.58} & 0.86/0.76/0.69 & --/--/-- \\
		R2GenRL* & 0.85/0.67/0.62 & 0.77/0.48/0.41 & 0.84/0.65/0.60 & --/--/-- \\
		KMVE & 0.88/0.76/\textbf{0.74} & 0.82/0.64/0.57 & 0.87/0.76/0.73 & 3.50/3.80/1.91 \\
		\midrule
		InternVL-2.5(2B) & 0.89/0.76/0.73 & 0.83/0.66/0.57 & \textbf{0.88}/0.80/0.75 & 3.73/4.48/2.09 \\
		InternVL-2.5(4B) & 0.89/0.76/0.73 & 0.82/0.65/0.58 & 0.88/0.79/\textbf{0.75} & 3.64/4.38/2.14 \\
		Qwen2-VL(2B) & \textbf{0.89}/0.76/0.73 & \textbf{0.83}/0.64/0.57 & 0.88/0.79/0.74 & \textbf{4.04}/4.35/2.12 \\
		LLaVA-Med(7B) & 0.89/\textbf{0.78}/0.72 & 0.82/\textbf{0.67}/0.57 & 0.87/\textbf{0.80}/0.74 & 3.64/\textbf{4.50}/\textbf{2.17} \\
		HuatuoGPT\ldots(7B)& 0.86/0.77/0.71 & 0.79/0.64/0.55 & 0.85/0.78/0.72 & 2.95/4.10/2.15 \\
		\bottomrule
	\end{tabular}
\end{table}

MLLMs outperform traditional methods like DeltaNet~\cite{DeltaNet} and R2GenRL~\cite{R2GenRL}, as well as KMVE~\cite{KMVE_us_report}, across most metrics (Table~\ref{tab:combined_metrics_table}). For example, Qwen2-VL(2B) achieves a CIDEr score of 4.04 for Liver (+15.4\% over KMVE), and LLaVA-Med(7B) achieves 4.50 for Mammary (+18.4\%). Improvements in BLEU-4 and ROUGE-L are smaller but consistent, generally within 2–5\%. These gains highlight the ability of MLLMs to generate diverse, semantically rich, and clinically relevant reports compared to traditional methods, which rely on simpler, more repetitive language.

Among MLLMs, larger models like LLaVA-Med(7B) show slight advantages in some tasks, such as Mammary BLEU-4 (0.67 vs. 0.64 for Qwen2-VL(2B)) and Thyroid CIDEr (2.17 vs. 2.12). However, smaller models, like InternVL-2.5(2B), remain competitive, achieving a CIDEr score of 4.48 for Mammary, outperforming HuatuoGPT-Vision(7B) (4.10). Notably, these MLLMs have not yet been applied with the proposed fragment-based multilingual training pipeline, indicating potential for further improvements. The larger vocabularies of MLLMs (30k–150k tokens) allow for more nuanced outputs, avoiding the repetitive phrases seen in traditional models.

\subsection{Multilingual Training and Ablation Studies}
\begin{table}[ht]
	\caption{Overall Performance comparison of the previous best method KMVE~\cite{KMVE_us_report}, baselines, and the proposed multilingual fragment-based training ("Multi") approaches.}
	\centering
	\label{tab:multilingual_res}
	\begin{tabular}{l|ccccccc}
		\toprule
		\textbf{Experiment} & B1 & B4 & RL & C & BE-ZhF & BE-MF & MKF1 \\
		\midrule
		KMVE~\cite{KMVE_us_report} & 0.786 & 0.668 & 0.774 & 3.499 & 0.922 & 0.921 & 0.924 \\
		\midrule
		InternVL-2.5(2B) Vanilla & 0.789 & 0.679 & 0.797 & 3.970 & 0.931 & 0.930 & 0.934 \\
		InternVL-2.5(2B) Lora & 0.759 & 0.643 & 0.762 & 3.445 & 0.919 & 0.919 & 0.915 \\
		InternVL-2.5(4B) Vanilla & 0.789 & 0.674 & 0.793 & 3.910 & 0.930 & 0.930 & 0.933 \\
		InternVL-2.5(4B) Lora & 0.752 & 0.638 & 0.765 & 3.494 & 0.921 & 0.921 & 0.918 \\
		HuatuoGPT-Vision(7B) Vanilla & 0.775 & 0.652 & 0.772 & 3.636 & 0.924 & 0.923 & 0.926 \\
		HuatuoGPT-Vision(7B) Lora & 0.772 & 0.646 & 0.763 & 3.428 & 0.920 & 0.919 & 0.920 \\
		\midrule
		{\bfseries Qwen2-VL(2B) Multi} & 0.794 & 0.681 & 0.799 & 4.059 & 0.933 & 0.932 & 0.937 \\
		Qwen2-VL(2B) Vanilla & 0.787 & 0.673 & 0.790 & 3.963 & 0.929 & 0.928 & 0.934 \\
		Qwen2-VL(2B) Lora & 0.748 & 0.630 & 0.752 & 3.263 & 0.915 & 0.915 & 0.906 \\
		Qwen2-VL(2B) En2Zh & 0.792 & 0.679 & 0.798 & 4.073 & 0.932 & 0.932 & 0.936 \\
		Qwen2-VL(2B) Direct & 0.797 & 0.687 & 0.802 & 4.072 & 0.933 & 0.933 & 0.939 \\
		{\bfseries LLaVA-Med(7B) Multi} & 0.798 & {\bfseries 0.689} & {\bfseries 0.804} & {\bfseries 4.123} & {\bfseries 0.934} & {\bfseries 0.933} & 0.939 \\
		LLaVA-Med(7B) Vanilla & 0.791 & 0.680 & 0.796 & 3.980 & 0.931 & 0.930 & 0.936 \\
		LLaVA-Med(7B) Lora & 0.755 & 0.637 & 0.760 & 3.503 & 0.919 & 0.919 & 0.919 \\
		LLaVA-Med(7B) En2Zh & 0.797 & 0.686 & 0.802 & 4.108 & 0.934 & 0.933 & {\bfseries 0.940} \\
		LLaVA-Med(7B) Direct & {\bfseries 0.801} & 0.687 & 0.802 & 4.085 & 0.933 & 0.933 & 0.936 \\
		\midrule
		\midrule
		{\bfseries Qwen2-VL(2B) En Multi} & 0.773 & 0.648 & 0.753 & {\bfseries 3.768} & 0.908 & 0.921 & -- \\
		Qwen2-VL(2B) En Only & {\bfseries 0.779} & {\bfseries 0.656} & 0.752 & 3.720 & 0.906 & 0.919 & -- \\
		Qwen2-VL(2B) En Direct & 0.702 & 0.484 & 0.662 & 3.039 & 0.880 & 0.900 & -- \\
		{\bfseries LLaVA-Med(7B) En Multi} & 0.777 & 0.655 & {\bfseries 0.759} & 3.744 & {\bfseries 0.909} & {\bfseries 0.923} & -- \\
		LLaVA-Med(7B) En Only & 0.762 & 0.638 & 0.744 & 3.677 & 0.904 & 0.918 & -- \\
		LLaVA-Med(7B) En Direct & 0.706 & 0.475 & 0.653 & 2.908 & 0.876 & 0.896 & -- \\
		\bottomrule
	\end{tabular}
\end{table}

Table ~\ref{tab:multilingual_res} compares KMVE~\cite{KMVE_us_report}, baselines, and the proposed approaches. The baselines include "Vanilla," where ViT unfreezing + LoRA is applied, and "LoRA," which unfreezes only the projector and LoRA. The proposed "Multi" method integrates ViT unfreezing, LoRA, and multilingual fragment-based training using cross-language prompts. Additional configurations include "En2Zh," which generates Chinese reports from English queries, and "Direct," which is trained on directly translated datasets. While "En Direct" uses raw translations, other "En" configurations employ fragment-based translations as the testset.

The "Multi" configuration achieves the best results across all metrics, demonstrating its robustness and versatility. For example, LLaVA-Med(7B) "Multi" achieves a BLEU-4 score of 0.689 (+8.2\% over LoRA), a CIDEr of 4.123 (+17.7\%), and an MKF1 of 0.939 (+2.2\%). Against KMVE, "Multi" improves BLEU-4 by 3.1\% (0.689 vs. 0.668), CIDEr by 17.8\% (4.123 vs. 3.499), ROUGE-L by 3.9\% (0.804 vs. 0.774), BERTScore-ZhF1 by 1.3\% (0.934 vs. 0.922), and MKF1 by 1.6\% (0.939 vs. 0.924). Similarly, Qwen2-VL(2B) "Multi" achieves a BLEU-4 of 0.681 (+6.6\% over LoRA), a CIDEr of 4.059 (+24.4\%), and an MKF1 of 0.937 (+2.4\%).

For Chinese tasks, "Direct" training improves raw Chinese predictions but produces inconsistent English outputs, demonstrating that using roughly translated data can still enhance the performance of the original language. In contrast, "En2Zh" and "Multi" configurations achieve competitive results. For example, Qwen2-VL(2B) achieves a BLEU-4 of 0.679, a CIDEr of 4.073, and an MKF1 of 0.936 in "En2Zh," closely matching "Multi" (0.681, 4.059, and 0.937, respectively). Similarly, LLaVA-Med(7B) achieves a BLEU-4 of 0.686, CIDEr of 4.108, and MKF1 of 0.940 in "En2Zh," nearly identical to "Multi."

For English tasks, "En Multi" outperforms "En Only" (trained on fragment-translated English data only). For example, Qwen2-VL(2B) achieves a CIDEr of 3.768, and an MKF1 of 0.921 in "En Multi," compared to 3.720 and 0.919 in "En Only." Similarly, LLaVA-Med(7B) achieves a BLEU-4 of 0.655, a CIDEr of 3.744, and an MKF1 of 0.923 in "En Multi," compared to 0.638, 3.677, and 0.918 in "En Only." These results highlight the effectiveness of multilingual fragment-based training in improving both semantic metrics and clinical keyword matching, compared to approaches that rely solely on single-language information.

\section{Discussion and Conclusion}
This study presents a novel framework leveraging MLLMs specifically for multilingual, multi-organ ultrasound report generation, an application that, to our knowledge, remains largely unexplored. By incorporating fragment-based multilingual training, the proposed method achieves state-of-the-art performance, surpassing the recent best KMVE~\cite{KMVE_us_report} and baseline MLLMs across multiple evaluation metrics. Unfreezing the ViT backbone significantly enhances text-image alignment, resulting in substantial improvements in CIDEr scores of 15.3\% and 13.6\% for InternVL-2.5(2B) and LLaVA-Med(7B), respectively, compared to LoRA-only setups. Fragment-based multilingual training further improves performance, with gains of approximately 1.3\% in BLEU-4, 3.6\% in CIDEr, and 0.3\% in MKF1 over the Vanilla baseline, while achieving 3.1\% improvement in BLEU-4, 17.8\% in CIDEr, and 1.6\% in Matching Clinical Keywords F1 compared to KMVE.

The proposed approach directly generates native English reports, avoiding the need for post-hoc translation while reducing common errors such as extra information (highlighted in red), incorrect meanings (in brown), and missing content (strikethrough), as shown in Figure~\ref{fig:methods} section 7. Unlike previous methods, such as KMVE, which omits critical findings, or ViT unfreezing + LoRA alone, often introduce irrelevant or incorrect details, the proposed method ensures semantically accurate and contextually consistent outputs. To ensure clinical relevance, the translated fragments used in multilingual training were roughly examined by radiologists, minimizing the risk of major semantic errors during translation.

Despite its promising results, this study has several limitations that warrant further investigation. Ultrasound datasets remain scarce and fragmented compared to large-scale standardized datasets for X-ray or CT. While we use one of the largest available ultrasound datasets, its limited scale may impact generalizability to broader clinical settings and diverse populations. Additionally, our fragment-based approach depends on the statistical priors of the dataset, meaning deployment in settings with significantly different reporting patterns or terminology preferences may require additional fine-tuning. The translated fragments used in multilingual training were validated by radiologists to ensure clinical correctness, but variations in reporting conventions and stylistic nuances across languages or healthcare systems may not be fully captured, potentially affecting linguistic naturalness in real-world settings. Furthermore, this study focuses on three common ultrasound applications (mammary, thyroid, and liver), and extending the approach to other anatomical sites, such as cardiac or obstetric imaging, remains an open challenge. Finally, while our bilingual framework demonstrates strong performance for Chinese-English translation, extending it to additional languages with diverse medical terminologies and reporting conventions is an important direction for future research.

In conclusion, this work offers a scalable, innovative framework for multilingual US report generation, addressing key challenges in text-image alignment, semantic consistency, and language adaptability. By seamlessly integrating into existing machine-learning and deployment pipelines, the proposed approach sets a foundation for future advancements in medical AI. While challenges such as dataset size, dependency on statistical priors, and limited organ and language coverage remain, this work highlights critical directions for future research and contributes a significant step toward scalable, clinically relevant ultrasound report generation.

\bibliographystyle{splncs04}
\bibliography{refs.bib}

%Author, F.: Article title. Journal \textbf{2}(5), 99--110 (2016)
%Author, F., Author, S.: Title of a proceedings paper. In: Editor,
%F., Editor, S. (eds.) CONFERENCE 2016, LNCS, vol. 9999, pp. 1--13.
%Springer, Heidelberg (2016). \doi{10.10007/1234567890}
%Author, F., Author, S., Author, T.: Book title. 2nd edn. Publisher,
%Author, A.-B.: Contribution title. In: 9th International Proceedings
%on Proceedings, pp. 1--2. Publisher, Location (2010)
%LNCS Homepage, \url{http://www.springer.com/lncs}, last accessed 2023/10/25
%\end{thebibliography}
\end{document}